\documentclass[10pt, conference, compsocconf]{IEEEtran}

\usepackage{hyperref}

\ifCLASSINFOpdf
   \usepackage[pdftex]{graphicx}

\else

\fi

\usepackage{color}
\usepackage{listings}
\newenvironment{ppl}{\small\ttfamily}{}

\begin{document}
%
\title{A Dataflow Language for Decentralised Orchestration of Web Service Workflows}


\author{\IEEEauthorblockN{Ward Jaradat, Alan Dearle, and Adam Barker}
\IEEEauthorblockA{School of Computer Science, University of St Andrews, 
 North Haugh, St Andrews, Fife, KY16 9SX, United Kingdom\\
\{ward.jaradat, alan.dearle, adam.barker\}@st-andrews.ac.uk}
}


%


\maketitle

\begin{abstract}
Orchestrating centralised service-oriented workflows presents significant scalability challenges that include: the consumption of network bandwidth, degradation of performance, and single points of failure. This paper presents a high-level dataflow specification language that attempts to address these scalability challenges. This language provides simple abstractions for orchestrating large-scale web service workflows, and separates between the workflow logic and its execution. It is based on a data-driven model that permits parallelism to improve the workflow performance. We provide a decentralised architecture that allows the computation logic to be moved ``closer" to services involved in the workflow. This is achieved through partitioning the workflow specification into smaller fragments that may be sent to remote orchestration services for execution. The orchestration services rely on proxies that exploit connectivity to services in the workflow. These proxies perform service invocations and compositions on behalf of the orchestration services, and carry out data collection, retrieval, and mediation tasks. The evaluation of our architecture implementation concludes that our decentralised approach reduces the execution time of workflows, and scales accordingly with the increasing size of data sets.

\end{abstract}


%
\IEEEpeerreviewmaketitle

\section{Introduction}

Centralised orchestration is typically used to execute large-scale service-oriented workflows. This centralised approach provides control over the workflow, supports process automation, and permits the workflow logic to be encapsulated, modified, or extended as necessary at a central location. However, it presents a set of research challenges, these include: the consumption of network bandwidth, degradation of performance, and single points of failure [1]. These challenges are particularly prominent when dealing with highly distributed data-intensive workflows such as scientific workflows used in the area of astronomy [2]. These workflows are commonly composed as Directed Acyclic Graphs (DAG), and involve large quantities of intermediate data that need to be routed through a centralised engine. The specification of these workflows is complex as they consist of asynchronous, distributed and concurrent processes.\\

This paper aims to address these challenges by providing a simple high-level dataflow language that reduces the complexity of specifying service-oriented workflows. In our approach, we provide a decentralised orchestration architecture with which workflows specified using our language can be partitioned into smaller fragments that may be sent to distributed orchestration services. These orchestration services collaborate together to execute the workflow with each orchestration service executing part of the workflow specification. Our research hypothesis states that this approach reduces the overall communication between services, and avoids potential performance bottlenecks due to the distribution of data and the absence of a centralised orchestration engine.\\

The remainder of this paper is structured as follows: Section 2 presents an overview of our dataflow language. Section 3 presents the language design. Section 4 provides a set of examples that describe the language support for common dataflow patterns. Section 5 explains the data type system of the language. Section 6 presents our distributed architecture that is used to execute the language, and provides an evaluation of our architecture implementation. Section 7 discusses related work in the area of our research. Finally, Section 8 concludes the paper and states future work.

\section{Language Overview}

Our language provides abstractions that define a set of services and coordinate the dataflow between them. It separates the workflow logic from its execution, which permits the workflow architect to focus on specifying the workflow without knowledge of how it is executed. Unlike typical dataflow languages that separate computation and coordination, we believe that creating a dataflow language that combines both has much merit and deserves further investigation. This language is characterised as a simple dataflow language that supports parallelism, and provides a data-driven execution model. 

\subsection{Language Characteristics}


\subsubsection{Simplicity}
The language permits the workflow to be expressed in simple high-level abstractions that describe only the information required to perform service invocations, compose services together, and coordinate the dataflow between them. Consequently, the workflow architect does not have to deal with low-level details related to the execution mechanism of the language such as process creation, placement and management of computation, communication, and synchronisation.

\subsubsection{Parallelism}

The language provides implicit parallelism support to improve the execution performance of workflows. It permits the workflow specification to be represented as a Directed Acyclic Graph (DAG) that can be nested and may be decomposed into smaller graphs for concurrent execution. This provides ease of workflow partitioning and reduces the complexity of the language as it avoids the introduction of loops and control structures. In our approach, parallelism is automatically recognised by the language compiler. Furthermore, the result of any specified workflow will be the same at all times regardless of the execution of intricate parts of the workflow in parallel. Consequently, we can easily run and debug a workflow specified using our language on a single sequential machine and then compile it for execution on parallel machines.\\

\subsubsection{Data-driven execution model}
The language adopts a data-driven execution model that depends on the availability of data. For instance, a service invocation can only be executed when the input data that is required for its execution becomes available. This permits the data required for executing specific parts of the workflow to be obtained from multiple services.

\subsection{Simple Example}

In this section, we provide an example of a simple workflow that involves a number of services and combines common dataflow patterns used in complex scientific workflows. These patterns include the pipeline, data distribution and data aggregation patterns [2]. Figure 1 illustrates the structure of this workflow. In this example, \begin{ppl}a\end{ppl} represents the initial input data while the services are represented by \begin{ppl}S1\end{ppl}, \begin{ppl}S2\end{ppl}, \begin{ppl}S3\end{ppl}, \begin{ppl}S4\end{ppl}, \begin{ppl}S5\end{ppl}, and \begin{ppl}S6\end{ppl}. Some of these services produce intermediate output data that are used in the workflow such as \begin{ppl}x\end{ppl} and \begin{ppl}y\end{ppl}. The final workflow result is represented by \begin{ppl}z\end{ppl}.

\begin{figure}[!h]
\centerline{\includegraphics[scale=0.7]{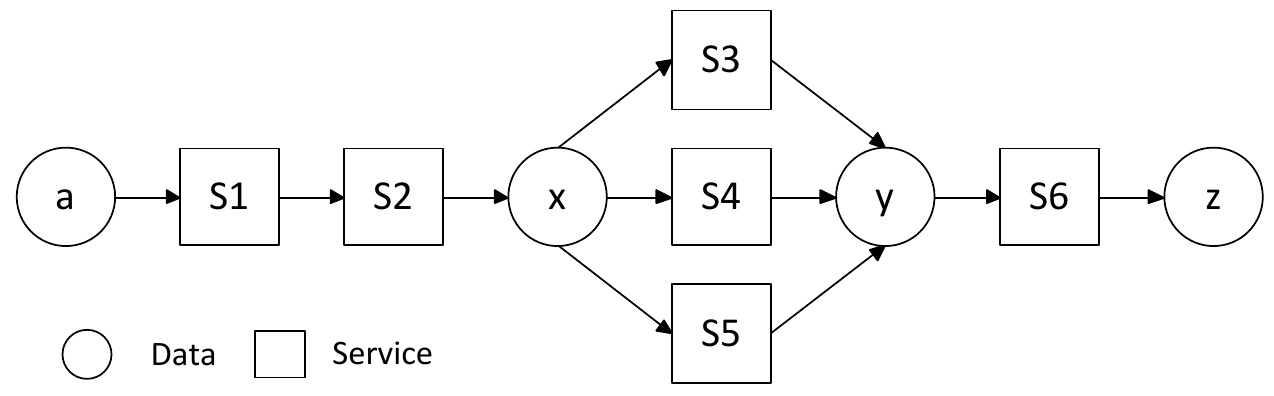}}
\label{fig:fig1}
\caption{Simple Workflow Example}
\end{figure}

Listing 1 shows the specification of this workflow using our language. This specification will be used throughout the paper to explain the language constructs, its support for common dataflow patterns, and data type system. Lines 1-13 define the services in the workflow and the service endpoints. Lines 15-18 provide the workflow interface that indicates the inputs and outputs of the workflow. Lines 20-31 consist of computation elements that represent service invocations, and coordination elements that indicate the direction of the dataflow between services. For instance, an input \begin{ppl}a\end{ppl} is used to invoke an operation \begin{ppl}Op1\end{ppl} provided by a service endpoint \begin{ppl}p1\end{ppl} in line 20. The invocation result is passed as input to another endpoint to create a service composition that produces the result \begin{ppl}x\end{ppl} through lines 21-22.\\

\begin{lstlisting}[keywords={description, service, port, is, input, output, int, any}, caption={Simple Workflow Specification}, frame=single]
01 description desc is http://ward.host.cs.
   st-andrews.ac.uk/documents/services.wsdl
02 service s1 is desc.Service1
...
07 service s6 is desc.Service6
08 port p1 is s1.Port1
...
13 port p6 is s6.Port6
14 
15 input: 
16    int a
17 output:
18    any x, y, z
19 
20 a -> p1.Op1
21 p1.Op1 -> p2.Op2
22 p2.Op2 -> x
23
24 x -> p3.Op3, p4.Op4, p5.Op5
25 p3.Op3 -> b
26 p4.Op4 -> c
27 p5.Op5 -> d
28
29 y = (b, c, d)
30 y -> p6.Op6
31 p6.Op6 -> z
\end{lstlisting}

\begin{figure*}[t]
\begin{lstlisting}[frame=single, caption={Language Grammar}]  
<specification> ::= <services>* <schema>* <interface> <dataflow>                 (Workflow Specification)

<services>      ::= <description> <service> <port>                                             (Services)
                
<description>   ::= description <name> is <URL>                                  (Description identifier)
<service>       ::= service <name> is <description-name> . <service-name>            (Service identifier)
<port>          ::= port <name> is <service-name> . <port-name>                         (Port identifier)

<schema>        ::= schema <name> is <URL>                                             (Data type schema)
<interface>     ::= <input> <output>                                                 (Workflow interface)
<input>         ::= input : <variables>                                                 (Defining Inputs)
<output>        ::= output : <variables>                                               (Defining Outputs)
<variables>     ::= <type> <name> [, <name>] *                                       (Defining Variables)
                                                                                     
<type>          ::= any | int | double | float | decimal                                (Base data types)
                |   byte | boolean | string | long | short    
                |   <schema> : <type-name>                                            (Complex data type)
                
<dataflow>      ::= <invocation>                                    (Service invocation without an input)
                |   <scalar> -> <invocation>          (Passing a scalar as input to a service invocation)
                |   <variable> -> <invocation>      (Passing a variable as input to a service invocation)
                |   <invocation> -> <variable>                   (Retrieving a service invocation output)
                |   <assignment>                                                             (Assignment)
                
<invocation>    ::= <port-name> . <operation-name> [. <parameter>]                   (Service invocation)
                
<assignment>    ::= <variable> = <scalar>                                           (Variable assignment)
                |   <variable> = <variable>
                |   <variable> = ( <variable> | <scalar> [, <variable> | <scalar> ] * )               
\end{lstlisting}
\label{fig:languagegrammar}

\end{figure*}

\section{Language Design}

The language grammar that is shown in Listing 2 provides a set of rules for defining services and constructing the dataflow between them. In this section we describe our language components that consist of identifiers, a workflow interface, and computation and coordination elements.

\subsection{Service Identifiers}

The language provides a set of identifiers that can be used to describe the services involved in a workflow and the service endpoints. These identifiers can be defined using the language constructs: \begin{ppl}description\end{ppl}, \begin{ppl}service\end{ppl}, and \begin{ppl}port\end{ppl}. The \begin{ppl}description\end{ppl} construct is used to define an identifier for a particular service description document. This identifier can be used to define the services involved in the workflow, and permit the language compiler to retrieve information about these services such as their endpoints, operations, and supported data types. The \begin{ppl}service\end{ppl} construct defines a service identifier to be used in the workflow, while the \begin{ppl}port\end{ppl} construct defines an identifier for the service endpoint.\\

Listing 3 provides an example that shows how to define services and their endpoints in the language. The \begin{ppl}description\end{ppl} construct is used to define an identifier \begin{ppl}desc\end{ppl} for a particular service description document. This document is based on the Web Service Description Language (WSDL) and can be located by a URL. The \begin{ppl}service\end{ppl} construct is used to define an identifier \begin{ppl}s1\end{ppl} for a particular service \begin{ppl}Service1\end{ppl} provided by the service description identifier, and the \begin{ppl}port\end{ppl} construct is used define an identifier \begin{ppl}p1\end{ppl} for \begin{ppl}Port1\end{ppl} that is provided by the service identifier.\\

\begin{lstlisting}[keywords={description, service, port, is, input, output, int, any}, caption={Service Identifiers}, frame=single]
01 description desc is http://ward.host.cs.st-
         andrews.ac.uk/documents/services.wsdl
02 service s1 is desc.Service1
03 port p1 is s1.Port1
\end{lstlisting}

\subsection{Workflow Interface}

The workflow interface is used to indicate the initial inputs that are required to execute the workflow, and the outputs to be obtained from the workflow execution. It permits the inputs and outputs of the workflow to be defined using simple and complex data types. Listing 4 provides a workflow interface that consists of an input \begin{ppl}a\end{ppl} of integer type, and outputs \begin{ppl}x\end{ppl}, \begin{ppl}y\end{ppl} and \begin{ppl}z\end{ppl} of an arbitrary data type \begin{ppl}any\end{ppl}.\\

\begin{lstlisting}[keywords={description, service, port, is, input, output, int, any}, caption={Workflow Interface}, frame=single]
01 input:
02    int a
03 output:
04    any x, y, z
\end{lstlisting}

\subsection{Computation and Coordination Elements}

The language permits the description of computation elements that represent service invocations, and the data passed to them. The output of a particular service invocation can be associated with an identifier, or passed directly to another service invocation to create a service composition. Each service invocation consists of a port identifier and an associated operation separated by a dot symbol. The dataflow specification language provides the coordination symbol \begin{ppl}->\end{ppl} to indicate the direction of the data passed to or retrieved from service invocations. The execution of a particular service invocation can take place when all the input parameters for that service invocation are satisfied during execution. This permits a service invocation to be executed as soon as the data that is required for its execution becomes available from different sources. \\

Listing 5 shows a simple service invocation where \begin{ppl}a\end{ppl} is used as an input to invoke an operation \begin{ppl}Op1\end{ppl} provided by a service endpoint \begin{ppl}p1\end{ppl}. The result of this service invocation is represented by \begin{ppl}b\end{ppl}.\\

\begin{lstlisting}[keywords={description, service, port, is, input, output, int, any}, caption={Simple Service Invocation Example}, frame=single]
01 a -> p1.Op1
02 p1.Op1 -> b
\end{lstlisting}

\section{Dataflow Patterns Support}

Our dataflow specification language provides support for common dataflow patterns that include: pipeline, data distribution and data aggregation patterns. These simple patterns can be used to create complex dataflows by combining them together, and are commonly found in scientific workflows [2].\\

\subsection{Pipeline Dataflow Pattern}

The pipeline pattern is used for chaining several services, where an output of one service provides the input of another service. Figure 2 shows this pattern where \begin{ppl}a\end{ppl} is used as an input for invoking service \begin{ppl}S1\end{ppl}. The service invocation output is then passed to service \begin{ppl}S2\end{ppl} that returns the final output \begin{ppl}x\end{ppl}. Listing 6 shows the specification of this pattern in our language, based on the workflow example presented earlier in section 2. In this example, \begin{ppl}a\end{ppl} is used as input to invoke a service endpoint \begin{ppl}p1.Op1\end{ppl} in line 20. The output of this service invocation is then passed to another service as input in line 21, and the final output is represented by \begin{ppl}x\end{ppl} in line 22.\\

\begin{figure}[!h]
\centerline{\includegraphics[scale=0.7]{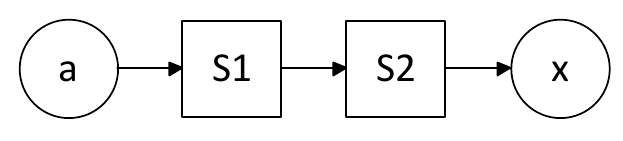}}
\label{fig:fig2}
\caption{Pipeline Dataflow Pattern}
\end{figure}

\begin{lstlisting}[keywords={description, service, port, is, input, output, int, any}, caption={Pipeline Dataflow Pattern Specification}, frame=single]
...
02 service s1 is desc.Service1
03 service s2 is desc.Service2
...
08 port p1 is s1.Port1
09 port p2 is s2.Port2
...
15 input: 
16    int a
17 output:
18    any x, y, z
19 
20 a -> p1.Op1
21 p1.Op1 -> p2.Op2
22 p2.Op2 -> x
\end{lstlisting}

\subsection{Data Distribution Pattern}

The data distribution pattern is used for distributing data to multiple services for processing. Figure 3 illustrates this pattern where \begin{ppl}x\end{ppl} represents the output of service \begin{ppl}S2\end{ppl}, which is used as input to invoke the services: \begin{ppl}S3\end{ppl}, \begin{ppl}S4\end{ppl}, and \begin{ppl}S5\end{ppl}. Listing 7 shows the specification of this pattern where \begin{ppl}x\end{ppl} is used to invoke \begin{ppl}p3.Op3\end{ppl}, \begin{ppl}p4.Op4\end{ppl}, and \begin{ppl}p5.Op5\end{ppl} in line 24. This finite sequence of service invocations represents the simplest parallel data structure in our language where each service invocation is executed concurrently. The invocation results are \begin{ppl}b\end{ppl}, \begin{ppl}c\end{ppl}, and \begin{ppl}d\end{ppl} as shown in lines 25-27. \newpage

\begin{figure}[!h]
\centerline{\includegraphics[scale=0.7]{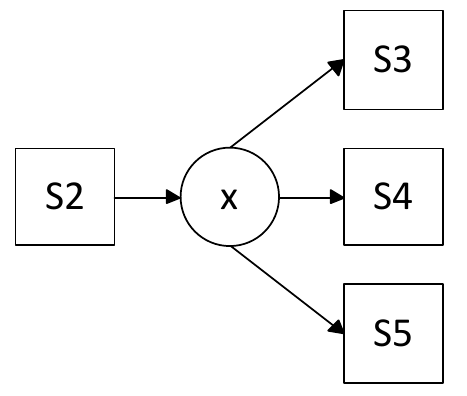}}
\label{fig:fig3}
\caption{Data Distribution Pattern}
\end{figure}

\begin{lstlisting}[keywords={description, service, port, is, input, output, int, any}, caption={Data Distribution Pattern Specification}, frame=single]
...
04 service s3 is desc.Service3
05 service s4 is desc.Service4   
06 service s5 is desc.Service5
...
10 port p3 is s3.Port3
11 port p4 is s4.Port4
12 port p5 is s5.Port5
...
15 input: 
16    int a
17 output:
18    any x, y, z
...
24 x -> p3.Op3, p4.Op4, p5.Op5
25 p3.Op3 -> b
26 p4.Op4 -> c
27 p5.Op5 -> d
\end{lstlisting}

\subsection{Data Aggregation Pattern}

The data aggregation pattern is used for aggregating data obtained from multiple services and sending it to a service that acts as a data sink. In this section we present a couple of techniques for specifying this dataflow pattern in our language. The first technique uses a tuple to combine data for invoking a particular service, whereas the second technique uses data routing to forward the data obtained from multiple services to a particular service directly.\\

\subsubsection{Data aggregation technique using a tuple}

Figure 4 illustrates this pattern where the outputs of \begin{ppl}S3\end{ppl}, \begin{ppl}S4\end{ppl}, and \begin{ppl}S5\end{ppl} are combined in a tuple \begin{ppl}y\end{ppl} that is passed to \begin{ppl}S6\end{ppl}, which produces an output \begin{ppl}z\end{ppl}. Listing 8 shows the specification of this pattern in which the service invocation results \begin{ppl}b\end{ppl}, \begin{ppl}c\end{ppl}, and \begin{ppl}d\end{ppl} are combined in a tuple that is used to invoke \begin{ppl}p6.Op6\end{ppl} in lines 29-30. The final output is represented by \begin{ppl}z\end{ppl} in line 31.

\begin{figure}[!h]
\centerline{\includegraphics[scale=0.7]{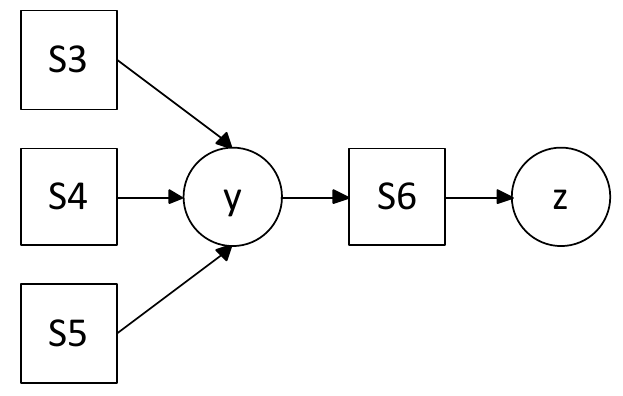}}
\label{fig:fig4}
\caption{Data Aggregation Pattern Using a Tuple}
\end{figure}

\begin{lstlisting}[keywords={description, service, port, is, input, output, int, any}, caption={Data Aggregation Pattern Specification Using a Tuple}, frame=single]
...
07 service s6 is desc.Service6
...
13 port p6 is s6.Port6
14 
15 input: 
16    int a
17 output:
18    any x, y, z
...
29 y = (b, c, d)
30 y -> p6.Op6
31 p6.Op6 -> z
\end{lstlisting}

\subsubsection{Data aggregation technique using data routing}

Figure 5 illustrates passing the outputs of services \begin{ppl}S3\end{ppl}, \begin{ppl}S4\end{ppl}, and \begin{ppl}S5\end{ppl} directly to service \begin{ppl}S6\end{ppl}. Listing 9 shows the specification of this pattern where the invocation outputs of \begin{ppl}p3.Op3\end{ppl}, \begin{ppl}p4.Op4\end{ppl}, and \begin{ppl}p5.Op5\end{ppl} are passed directly as input parameters \begin{ppl}a\end{ppl}, \begin{ppl}b\end{ppl}, and \begin{ppl}c\end{ppl} to invoke \begin{ppl}p6.Op6\end{ppl} that produces \begin{ppl}z\end{ppl} in lines 25-28.

\begin{figure}[!h]
\centerline{\includegraphics[scale=0.7]{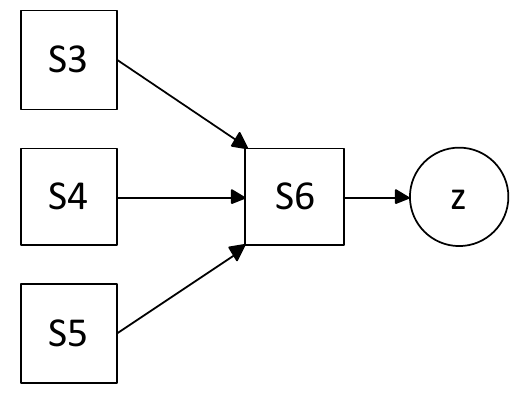}}
\label{fig:fig5}
\caption{Data Aggregation Pattern Using Data Routing}
\end{figure}

\begin{lstlisting}[keywords={description, service, port, is, input, output, int, any}, caption={Data Aggregation Pattern Specification Using Data Routing}, frame=single]
...
07 service s6 is desc.Service6
...
13 port p6 is s6.Port6
14 
15 input: 
16    int a
17 output:
18    any x, y, z
...
25 p3.Op3 -> p6.Op6.a
26 p4.Op4 -> p6.Op6.b
27 p5.Op5 -> p6.Op6.c
28 p6.Op6 -> z
\end{lstlisting}

\section{Data Type System}

\subsection{Base Data Types and Scalars}

In our language, we provided a set of base types that are expressed in a type system matching those defined in the XML Schema standard [3]. The base data types supported by our language include: \begin{ppl}byte\end{ppl}, \begin{ppl}boolean\end{ppl}, \begin{ppl}string\end{ppl}, \begin{ppl}int\end{ppl}, \begin{ppl}float\end{ppl}, \begin{ppl}double\end{ppl}, \begin{ppl}decimal\end{ppl}, \begin{ppl}long\end{ppl}, and \begin{ppl}short\end{ppl}. In order to represent arbitrary data types, we provided support for the union type \begin{ppl}any\end{ppl}. These base data types are used to define the inputs and outputs types in the workflow interface. The language supports scalars of base data types, and scalar values can be assigned to variables or used as input parameters for service invocations. It provides single-assignment for variables, and the variable type may be inferred from the scalar value being assigned to that variable. If a scalar is used to invoke a service endpoint, the language analyser performs data type checking to verify that the service endpoint accepts an input whose data type matches that of the scalar.

\subsection{Complex Data Types}

The language supports complex data types that can be obtained from external data type schemas based on XML [3]. Listing 10 provides an example that defines an external data type schema identifier \begin{ppl}schm\end{ppl} using the \begin{ppl}schema\end{ppl} construct as shown in line 1. This identifier is used to define an input \begin{ppl}x\end{ppl} of the data type \begin{ppl}newType\end{ppl} in line 11. We have shown how to create a complex data type variable represented by a tuple that can be assigned several values in Listing 8. Similarly, the input \begin{ppl}x\end{ppl} may provide parameters that can be assigned with values that permit it to be used in the workflow. \\

\begin{lstlisting}[keywords={description, service, port, is, input, output, int, any, schema}, caption={External Data Type Schema Definition}, frame=single]
01 schema schm is http://ward.host.cs.st-andrews.
   ac.uk/documents/types.xsd
...
10 input:
11    schm:newType x
12 output:
13    any z
...
\end{lstlisting}

\section{Implementation and Evaluation}

\subsection{Decentralised Orchestration Architecture}

We have implemented a compiler in order to execute workflows based on our language. This compiler is built from a set of procedures matching the production rules of the language grammar, and uses a recursive descent parser and analyser to ensure the correctness of the workflow specification. It generates data structures that form the workflow graph in which the nodes may represent service invocations, with arcs as data dependencies. Typically, graphs are used to guide the computation process where the dataflow tokens represent inputs for service invocations. These invocations may become active and ready for execution when all the required input parameters become available. Upon the execution of a service invocation, its output value is passed to other service invocations that may depend on it, which themselves may become activated. This process continues until there are no more active invocations. \\

In order to improve the workflow performance we have built a decentralised orchestration architecture, which permits the workflow specification to be partitioned into smaller fragments for execution at remote locations ``closer" to the services and resources involved in the workflow. Figure 6 shows our architecture that consists of distributed orchestration services and proxies, which collaborate together to execute a particular workflow. The workflow specification can be analysed and compiled by an orchestration service, and partitioned into smaller fragments that may be transmitted to remote orchestration services. These orchestration services rely on proxies to exploit connectivity to services in the workflow. These proxies perform service invocations and compositions on behalf of the orchestration services, and carry out data collection, retrieval, and mediation tasks.

\begin{figure}[!h]
\centerline{\includegraphics[scale=0.9]{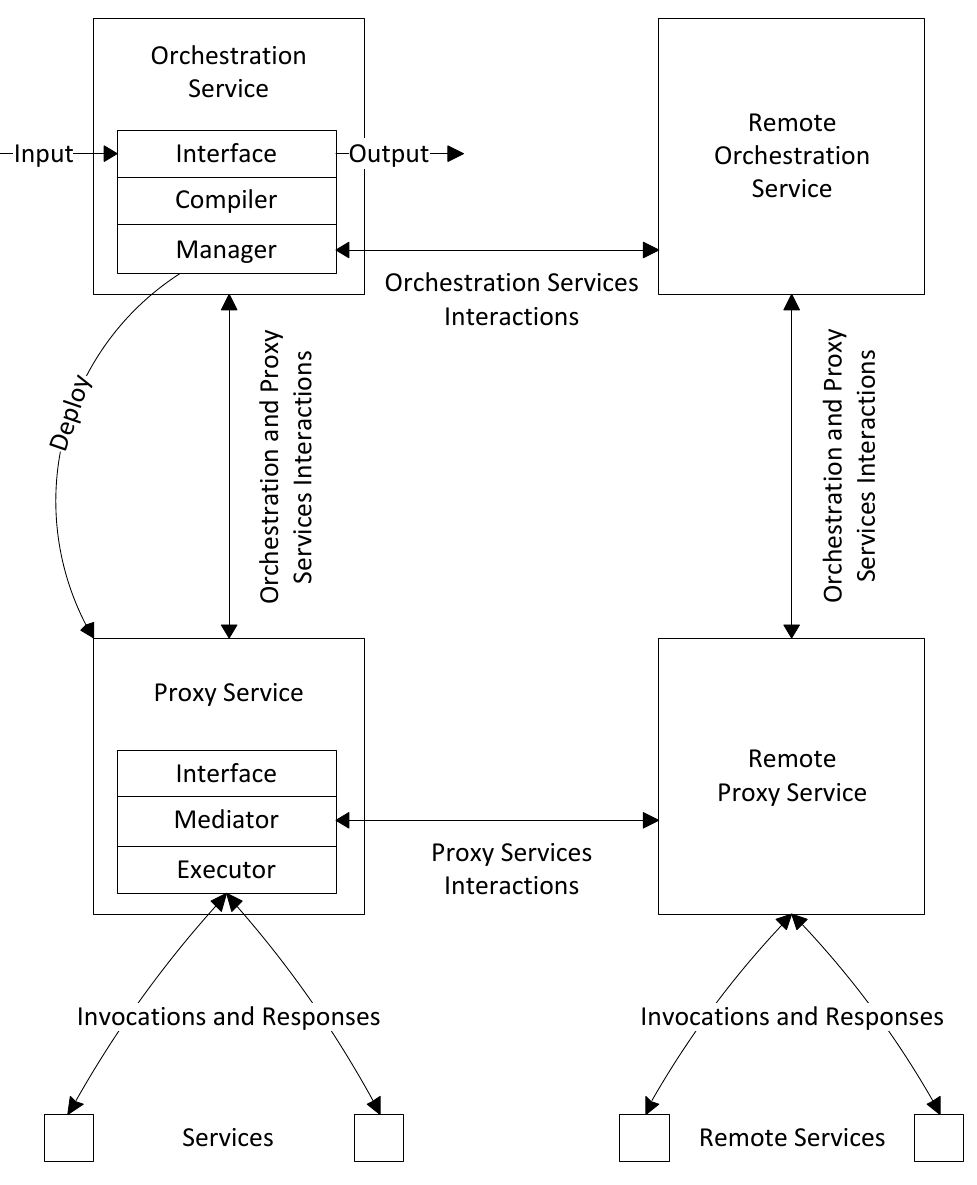}}
\label{fig:fig6}
\caption{Decentralised Orchestration Architecture}
\end{figure}

\subsection{Description and Configuration of Experiments}

We have conducted a number of experiments that aim to evaluate the performance of our architecture during the execution of workflows. These experiments involved testing services that were deployed on Amazon EC2 regions. Firstly, we specified a workflow based on a pipeline dataflow pattern in which the output size of data increases with each service invocation, and each test service produces an output that is double the size of the input used to invoke it. Secondly, we specified a workflow based on a data aggregation pattern in which the final output size is double the size of the input data that was collected from multiple test services. Thirdly, we specified a workflow based on a data distribution pattern where an initial test service is invoked to produce an output that is sent to multiple test services, and each test service produces an output of approximately the same size of the received input. Each of these workflows was executed 20 times by our architecture in centralised and decentralised configurations. \\

Figure 7 illustrates the centralised configuration where a single orchestration service \begin{ppl}O\end{ppl} manages a proxy \begin{ppl}P\end{ppl}. The test services \begin{ppl}T1\end{ppl}, \begin{ppl}T2\end{ppl}, \begin{ppl}T3\end{ppl}, and \begin{ppl}T4\end{ppl} are used in each workflow. 

\begin{figure}[!h]
\centerline{\includegraphics[scale=0.9]{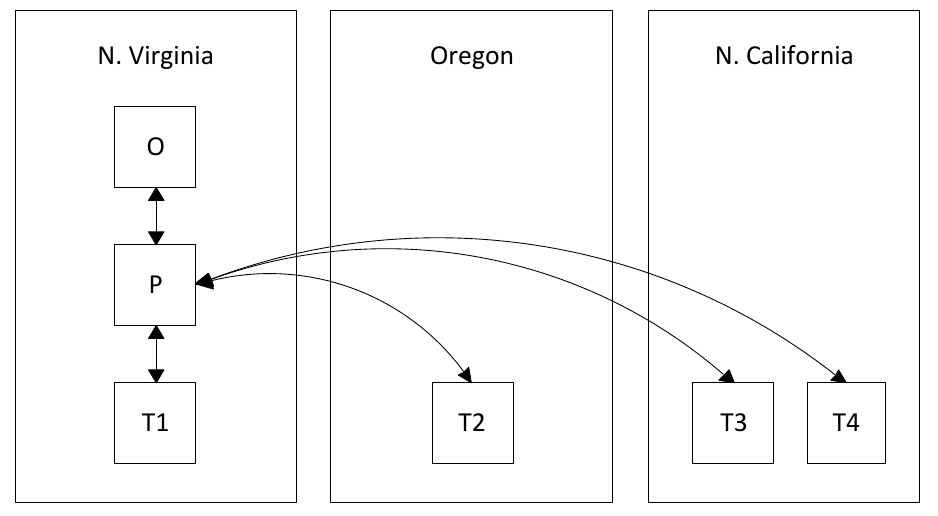}}
\label{fig:fig7}
\caption{Centralised Configuration}
\end{figure}

Figure 8 illustrates the decentralised configuration where multiple orchestration services \begin{ppl}O1\end{ppl}, \begin{ppl}O2\end{ppl} and \begin{ppl}O3\end{ppl} manage the proxies \begin{ppl}P1\end{ppl}, \begin{ppl}P2\end{ppl}, and \begin{ppl}P3\end{ppl} respectively.

\begin{figure}[!h]
\centerline{\includegraphics[scale=0.9]{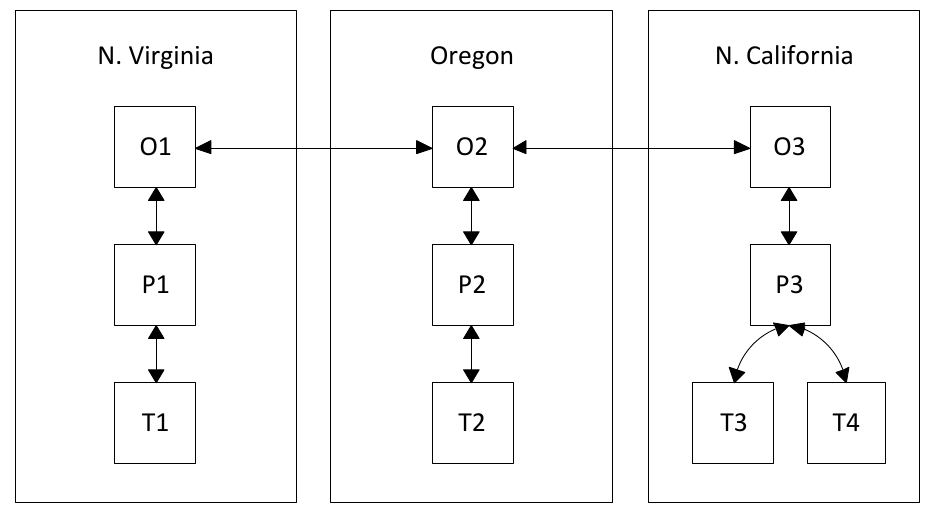}}
\label{fig:fig8}
\caption{Decentralised Configuration}
\end{figure}


\subsection{Performance Analysis}

\begin{figure}
\centerline{\includegraphics[scale=0.8]{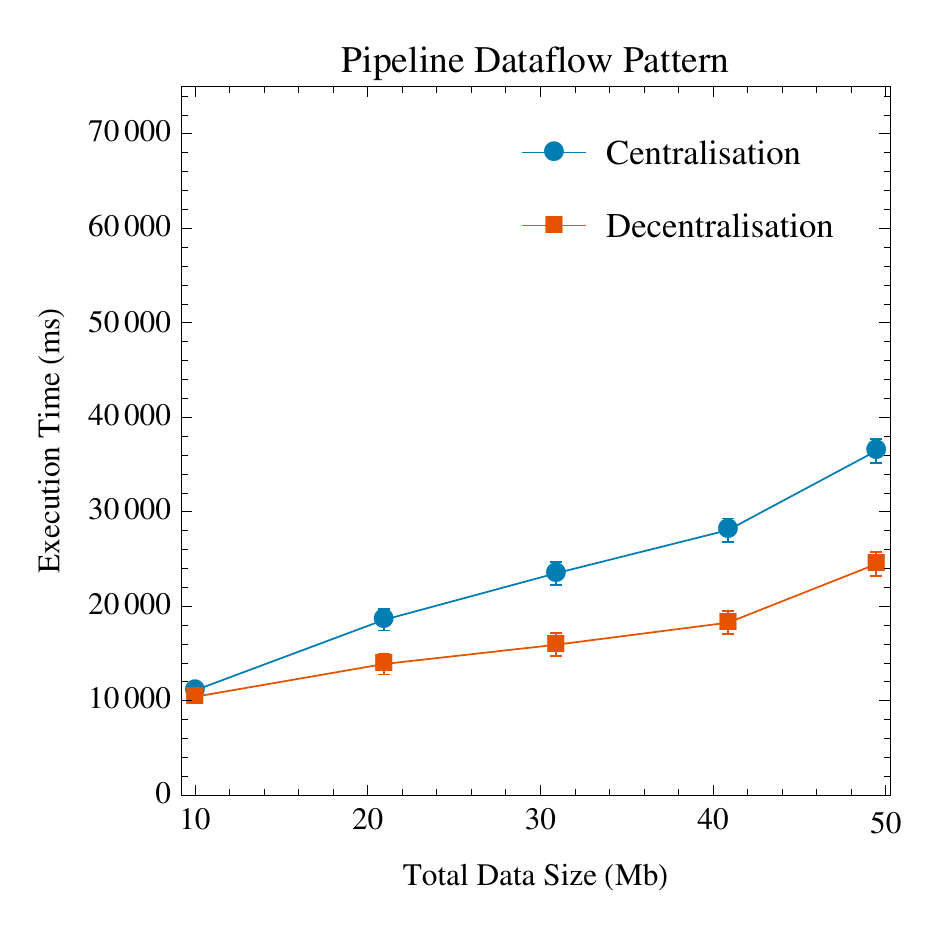}}
\centerline{\includegraphics[scale=0.8]{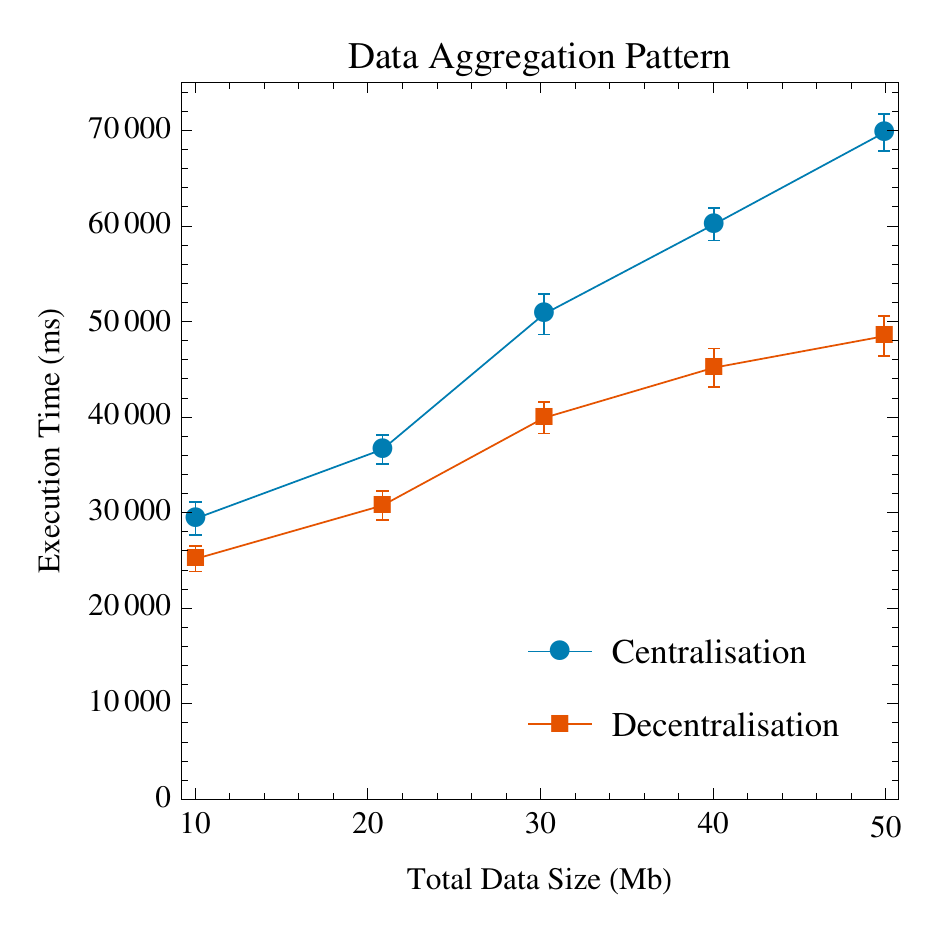}}
\centerline{\includegraphics[scale=0.8]{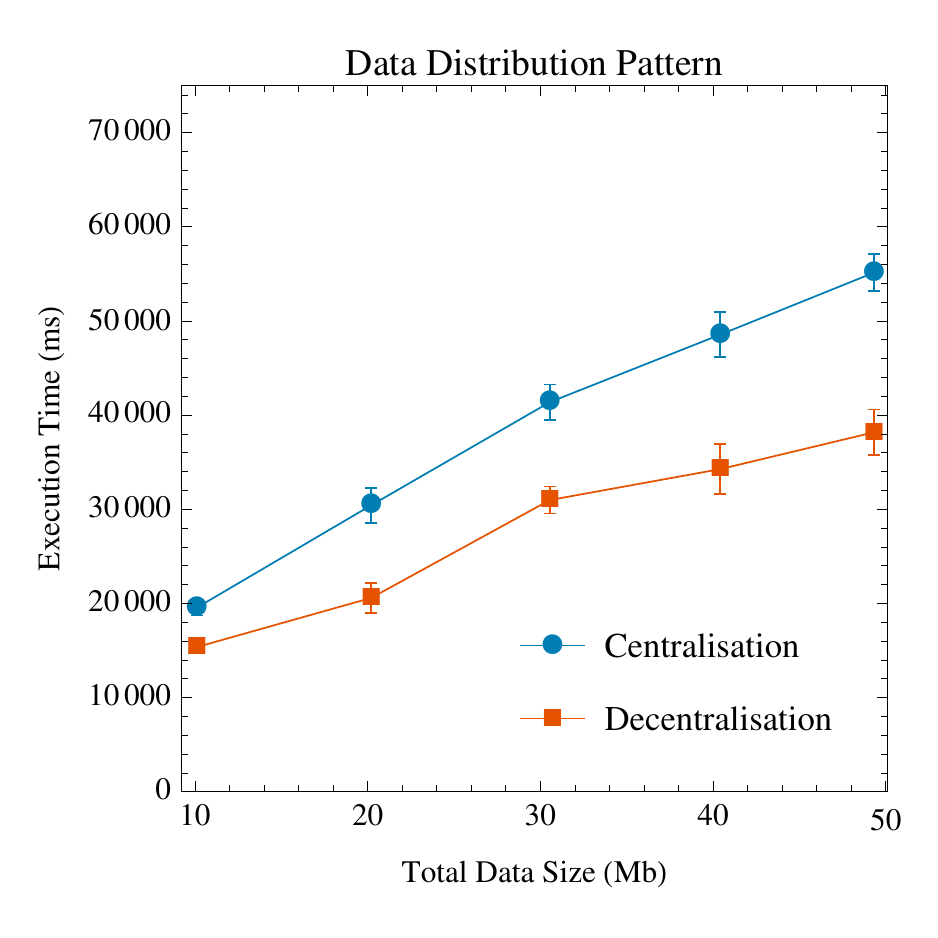}}
\label{fig:fig9}
\caption{Experimental Results}
\end{figure}

Table 1 provides the mean speedup rate for each experiment. It was calculated by dividing the mean time for executing the workflow using a single orchestration service, over the mean time for executing it using multiple orchestration services. The performance analysis verifies our research hypothesis, which states that decentralised orchestration reduces the overall execution time of the workflow compared with centralised orchestration. In order to explain our results, when using a centralised orchestration service, large intermediate copies of data pass through the orchestration service before they are sent across the network to other services in the workflow. This increases the workflow execution time and may overload the orchestration service. In our architecture, the intermediate data are stored by the proxies and used only when required during the workflow. This reduces the amount of data that is communicated between services and improves the workflow execution time. 

    \begin{table}[!h]\normalsize
    \caption{Mean Speedup Rates}
    \begin{center}
    \begin{tabular}{| l | l | l | l |}

    \hline
    Experiment & Mean Speedup Rate \\ \hline
    Pipeline & 1.37 \\ \hline
    Data Aggregation & 1.30 \\ \hline
    Data Distribution & 1.41 \\ \hline
    \end{tabular}
    \end{center}
    \end{table}
    
Figure 9 displays a set of graphs that provide the total size of data communicated in the workflow, and the workflow execution time for each experiment.

\section{Related Work}

The Business Process Execution Language (BPEL) [4] is an orchestration language that provides a component model for describing services as collaborating processes, which interact with each other based on a business protocol. However, the language limitation becomes apparent in large-scale workflows that require considerable effort to specify parallel activities and the dependencies between them. It has seen limited applications in scientific workflows.\\

The Abstract Grid Workflow Language (AGWL) [5] is a language based on XML that describes Grid workflow applications. It uses activities and provides constructs to define the control flow in these activities, and the dataflow between different activities. This permits explicit parallelism that increases the language sophistication. Our language provides implicit parallelism and avoids using control flow constructs to reduce the language complexity.\\

The Grid Services Flow Language (GSFL) [6] features an activity model that describes the workflow activities, and a composition model that describes the service interactions. However, the configuration of services may need to be altered before execution. In our approach, the workflow architect does not need to modify the service configurations.\\

Taverna is a workflow management system that uses the Simple Conceptual Unified Flow Language (SCUFL) [7] to define service interactions with control and data links. However, the language execution is based on a centralised model that suffers from performance bottlenecks. In our approach, highly distributed data-intensive workflows can be executed using our architecture in a decentralised fashion.

Swift [8] is a workflow scripting language that uses futures to enable parallel behaviour, and has a limited set of data types, operators, and built-in functions. It uses a data-driven model that is similar to our language model where a particular function can be executed when the required inputs for that function become available. However, Swift focuses on coordinating the execution of legacy applications coded in various programming languages rather than orchestrating the dataflow between web services involved in a workflow.\\

Dryad [9] is a distributed execution infrastructure for data-parallel programs. It combines computational vertices with communication channels to form a dataflow graph. However, these graphs are explicitly developed by the programmer. In our approach, these graphs are implicitly generated by the language compiler.\\

DAGMan (Directed Acyclic Graph Manager) [10] is a meta-scheduler for Condor. It is used for scheduling jobs within a graph in which the vertices are programs, and the arcs are program dependencies. However, it is not catered for orchestrating service-oriented workflows.\\

The Flow-based Infrastructure for Composing Autonomous Services (FICAS) [11] is a service composition infrastructure for executing distributed dataflows. It uses the Compositional Language for Autonomous Services (CLAS) to describe the behaviour of services. It can be compiled into a control sequence to be executed by the runtime environment. However, this approach is intrusive as the services must altered and wrapped with a special interface.

\section{Conclusion and Future Work}

This paper has presented our dataflow language that provides high-level abstractions for defining large-scale web service workflows. It reduces the complexity of specifying workflows and permits efficient use of parallelism. The language provides constructs for defining service identifiers, a workflow interface, and computation and coordination elements that permit the specification of common dataflow patterns. We have built a distributed architecture to execute workflow specifications based on our language. This architecture consists of orchestration services that can analyse a particular workflow specification, and may partition it to smaller fragments for execution at remote locations. These orchestration services rely on proxies to exploit connectivity to services involved in the workflow by invoking services and composing them together. The evaluation of our architecture concludes that our approach provides a scalable solution for reducing the amount of data transfer, and improving the workflow execution time.\\

Future work will include investigating several workflow partitioning and execution optimisation techniques, and providing execution policies to accommodate performance optimisation and resource utilisation requirements. In order to address security requirements, we will provide information policies to regulate the dissemination of confidential data in workflows. Clearly, there are other potential areas to consider such as the placement of computation, load balancing, and handling failures. Further information about our work is available at http://bigdata.cs.st-andrews.ac.uk/.

\end{document}